\documentclass[12pt]{article}
\usepackage{amsmath,amsfonts,amssymb}
\usepackage{graphicx}

\ifx\pdfoutput\undefined
     \relax
\else
     \usepackage{epstopdf}
\fi

\makeatletter
\def\numberbysection{\@addtoreset{equation}{section}
    \def\theequation{\thesection.\arabic{equation}}}
\makeatother

\numberbysection

\newcommand{\be}{\begin{eqnarray}}
\newcommand{\ee}{\end{eqnarray}}
\newcommand{\non}{\nonumber}
\newcommand{\tr}{\mathop{\rm tr}\nolimits}

\newcommand{\np}{\nonumber\\}

\begin{document}

\begin{titlepage}
\strut\hfill UMTG--255
\vspace{.5in}
\begin{center}

\LARGE The QCD spin chain $S$ matrix\\
\vspace{1in}
\large Changrim Ahn \footnote{
       Department of Physics, Ewha Womans University,
       Seoul 120-750, South Korea},
       Rafael I. Nepomechie \footnote{
       Physics Department, P.O. Box 248046, University of Miami,
       Coral Gables, FL 33124 USA}
   and Junji Suzuki \footnote{
       Department of Physics, Faculty of Science, Shizuoka University,
       Ohya 836, Shizuoka, Japan}\\

\end{center}

\vspace{.5in}

\begin{abstract}
    Beisert {\it et al.} have identified an integrable $SU(2,2)$ quantum spin chain
    which gives the one-loop anomalous dimensions of certain operators
    in large $N_{c}$ QCD. We derive a set of nonlinear integral
    equations (NLIEs) for this model, and compute the scattering matrix of
    the various (in particular, magnon) excitations.
\end{abstract}

\end{titlepage}

\setcounter{footnote}{0}

\section{Introduction}\label{sec:intro}

The search for integrability in QCD has a long history (see e.g.
\cite{tH}-\cite{BBGK} and references therein). A remarkable recent
development is the discovery \cite{FHZ} that the one-loop mixing
matrix \footnote{Given a set of operators ${\cal O}^{M}(x)$, the
mixing matrix is defined by $\Gamma=Z^{-1}\cdot dZ/d\ln \Lambda$, where
$Z$ is the renormalization factor which makes correlation functions
of ${\cal O}^{M}_{ren}(x)=Z^{M}_{\ N}{\cal O}^{N}(x)$ finite, and
$\Lambda$ is the ultraviolet cutoff. See also \cite{MZ}.}
for the chiral gauge-invariant operators
\be
\tr f_{\alpha_{1}\beta_{1}}(x)\ldots f_{\alpha_{L}\beta_{L}}(x)
\label{chiralops}
\ee
in the limit $N_{c}\rightarrow \infty$
is given by the integrable spin-1 antiferromagnetic XXX Hamiltonian
\cite{ZF, KRS},
\be
\Gamma = {\alpha_{s} N_{c}\over 2 \pi}\sum_{l=1}^{L}\left[\frac{7}{6}
+ \frac{1}{2} \vec S_{l} \cdot \vec S_{l+1}
- \frac{1}{2}(\vec S_{l} \cdot \vec S_{l+1})^{2} \right] \,.
\label{Gammachiralops}
\ee
Here $f_{\alpha\beta}$ are the selfdual components of the Yang-Mills
field strength $F_{\mu
\nu}=\partial_{\mu}A_{\nu}-\partial_{\nu}A_{\mu} - ig_{YM} [A_{\mu}\,,
A_{\nu}]$ (where the gauge fields $A_{\mu}(x)$ are $N_{c} \times N_{c}$
Hermitian matrices), which together with the anti-selfdual components
$\bar f_{\dot\alpha\dot\beta}$ are defined by
\be
F_{\mu \nu}= \sigma^{\ \ \ \alpha\beta}_{\mu\nu}\, f_{\alpha\beta}
+\bar \sigma^{\ \ \ \dot\alpha\dot\beta}_{\mu\nu}\, \bar
f_{\dot\alpha\dot\beta}\,,
\ee
where $\sigma_{\mu\nu}=i \sigma_{2}(\sigma_{\mu}\bar
\sigma_{\nu}-\sigma_{\nu}\bar \sigma_{\mu})/4$,
$\bar \sigma_{\mu\nu}=-i (\bar\sigma_{\mu} \sigma_{\nu}
-\bar\sigma_{\nu} \sigma_{\mu})\sigma_{2}/4$ and
$\sigma_{\mu}=(1\,, \vec \sigma)$, $\bar\sigma_{\mu}=(1\,, -\vec \sigma)$.
Moreover, $\alpha_{s}=g_{YM}^{2}/4\pi$,
$\alpha_{s} N_{c}$ is the `t Hooft coupling \cite{tH} which is assumed to be small,
and $\vec S$ are spin-1 generators of $SU(2)$. Indeed, since $f_{\alpha\beta}$
has three independent components
\be
f_{+}=f_{11}\,, \quad f_{0}=\frac{1}{\sqrt{2}}(f_{12}+f_{21})\,, \quad
f_{-}=f_{22}\,,
\ee
the operators (\ref{chiralops}) can be identified with the Hilbert
space of a periodic spin-1 quantum spin chain of length $L$.  The
eigenvectors and eigenvalues of $\Gamma$, i.e. the linear combinations
of the operators (\ref{chiralops}) which are multiplicatively
renormalizable and their anomalous dimensions, respectively, can
therefore be obtained using the Bethe Ansatz \cite{Ta, Ba}.  In
particular, the anomalous dimensions are given by
\be
\gamma = {\alpha_{s} N_{c}\over 2 \pi}\left( \frac{7L}{6} -
\sum_{j=1}^{M_{l}}{2\over l_{j}^{2}+1} \right)\,,
\label{gamma}
\ee
where $\{ l_{1}\,, \ldots \,, l_{M_{l}} \}$ are roots of the Bethe
Ansatz equations (BAEs) \footnote{There is an additional
(zero-momentum) equation due to the cyclicity of the trace in the
operators.}
\be
\left(l_{j}+i\over l_{j}-i\right)^{L} = \prod_{k=1 \atop k\ne j}^{M_{l}}
{l_{j}-l_{k}+i \over l_{j}-l_{k}-i} \,.
\label{BAE}
\ee

This result was generalized in \cite{BFHZ} to gauge-invariant
operators with derivatives
\be
\tr\, (D^{m_{1}}f)\ldots (D^{m_{L}}f) \,,
\label{chiralopsgen}
\ee
where
\be
D^{m}f = D_{\alpha_{1}\dot \alpha_{1}} \ldots D_{\alpha_{m}\dot
\alpha_{m}} f_{\beta\gamma} + symmetrized
\ee
(complete symmetrization in the undotted and dotted indices,
respectively), and $D_{\mu} = \sigma_{\mu}^{\alpha\dot \alpha}
D_{\alpha\dot \alpha}$ is the usual Yang-Mills covariant derivative.
Namely, the one-loop mixing matrix for the operators
(\ref{chiralopsgen}) is given by an integrable $SO(4,2)=SU(2,2)$
(non-compact!)  quantum spin chain Hamiltonian with spins in the
representation with Dynkin labels [2,-3,0].  The anomalous dimensions
are given by
\be
\gamma = {\alpha_{s} N_{c}\over 2 \pi}\left( \frac{7L}{6}
- \sum_{j=1}^{M_{l}}{2\over l_{j}^{2}+1}
+ \sum_{j=1}^{M_{u}}{3\over u_{j}^{2}+9/4}\right)\,,
\label{gammagen}
\ee
where the BAEs are now given by \footnotemark[\value{footnote}]
\be
\left(l_{j}+i\over l_{j}-i\right)^{L} &=&
\prod_{k=1 \atop k\ne j}^{M_{l}}
{l_{j}-l_{k}+i \over l_{j}-l_{k}-i}
\prod_{k=1}^{M_{u}}
{l_{j}-u_{k}-i/2 \over l_{j}-u_{k}+i/2}\,, \non \\
\left(u_{j}-3i/2\over u_{j}+3i/2\right)^{L} &=&
\prod_{k=1 \atop k\ne j}^{M_{u}}
{u_{j}-u_{k}+i \over u_{j}-u_{k}-i}
\prod_{k=1}^{M_{l}}
{u_{j}-l_{k}-i/2 \over u_{j}-l_{k}+i/2}
\prod_{k=1}^{M_{r}}
{u_{j}-r_{k}-i/2 \over u_{j}-r_{k}+i/2}\,, \label{BAEgen} \\
1&=&
\prod_{k=1 \atop k\ne j}^{M_{r}}
{r_{j}-r_{k}+i \over r_{j}-r_{k}-i}
\prod_{k=1}^{M_{u}}
{r_{j}-u_{k}-i/2 \over r_{j}-u_{k}+i/2}\,. \non
\ee
As noted by Beisert {et al.}, a $u$-root corresponds to adding a
covariant derivative $D_{1{\dot 1}}$; and an $l$-root and an $r$-root
flip a left-Lorentz-spin $1\to 2$ and a right-spin ${\dot 1}\to{\dot
2}$, respectively.
The scaling dimensions and $SU(2)_{L} \times SU(2)_{R}$
quantum numbers are given by
\be
D=2L+M_{u}\,, \quad S_{1}=L + {1\over 2}M_{u} - M_{l}\,, \quad
S_{2}= {1\over 2}M_{u} - M_{r}\,,
\label{quantumnos}
\ee
respectively.

As noted in \cite{BFHZ}, the BAEs (\ref{BAEgen}) can be obtained from
those of the ``beast'' form of ${\cal N}=4$ SYM \cite{BS} by
truncating the supergroup $SU(2,2|4)$ down to the Bosonic subgroup
$SU(2,2)$.  \footnote{For some early references on integrable
$gl(n|m)$ spin chains, see e.g. \cite{graded}.} Much attention has
been focused on the $S$ matrix of ${\cal N}=4$ SYM and of the
corresponding string theory (see e.g. \cite{Smatrix} ).

For the pure spin-1 problem (\ref{gamma}), (\ref{BAE}), the ground
state for large $L$ is described by a ``sea'' of approximate
``2-strings'' of $l$-roots \cite{Ta, Ba} (in contrast to the case of
the spin-1/2 antiferromagnetic XXX chain, for which the ground state
is described by a sea of {\it real} roots).  The excitations consist
of ``spinons'' (roughly speaking, ``holes'' in the sea) which carry
RSOS \cite{RSOS} quantum numbers.  The spinon-spinon $S$ matrix was
found by indirect methods in \cite{ABL, Re}, correcting the result
obtained in \cite{Ta} using the string hypothesis.
A nonlinear integral equation (NLIE) \cite{NLIE, NLIE2} has
been obtained for this model \cite{Su1}-\cite{Su2}, which does not
rely on the string hypothesis and provides a
more direct way to compute the $S$ matrix \cite{HRS}. The NLIE of the
$SU(2)$ sector of ${\cal N}=4$ SYM has been studied in \cite{FFGR}.

For the general case (\ref{gammagen}), (\ref{BAEgen}), the ground
state is still a sea of approximate 2-strings of $l$-roots, since the
$u$-roots contribute positively to the energy (and the $r$-roots do
not contribute at all).  Hence, there are again spinon excitations
corresponding to holes in the sea.  However, there are now also
``magnon'' excitations, corresponding to $u$-roots \cite{BFHZ}.

Our main objective here is to further investigate these magnon
excitations, and in particular, to compute the magnon-magnon $S$
matrix.  Owing to the nontrivial nature of the ground state, this $S$
matrix (like the spinon-spinon $S$ matrix) must be computed with care:
using the string hypothesis as in \cite{Ta} gives an incorrect result.
To this end, we first derive in Sec.  \ref{sec:NLIE} a set of NLIEs
for the model.  Although we do not invoke the string hypothesis, we do
make a certain analyticity assumption in order to describe the
$u$-roots. For simplicity, we restrict to real $u$-roots, and we do
not consider $r$-roots.
We then use these NLIEs to determine the energy and
momentum of the excitations (Sec.  \ref{sec:energymom}), and their $S$
matrices (Sec.  \ref{sec:Smatrix}).  We end in Sec.
\ref{sec:conclude} with a brief discussion of our results.

\section{Nonlinear integral equations}\label{sec:NLIE}

We restrict our attention to the case without $r$-roots
($M_{r}=0$), for which the BAEs (\ref{BAEgen}) reduce to
\be
\left(l_{j}+i\over l_{j}-i\right)^{L} &=&
\prod_{k=1 \atop k\ne j}^{M_{l}}
{l_{j}-l_{k}+i \over l_{j}-l_{k}-i}
\prod_{k=1}^{M_{u}}
{l_{j}-u_{k}-i/2 \over l_{j}-u_{k}+i/2}\,, \label{BAEgen1} \\
\left(u_{j}-3i/2\over u_{j}+3i/2\right)^{L} &=&
\prod_{k=1 \atop k\ne j}^{M_{u}}
{u_{j}-u_{k}+i \over u_{j}-u_{k}-i}
\prod_{k=1}^{M_{l}}
{u_{j}-l_{k}-i/2 \over u_{j}-l_{k}+i/2} \,. \label{BAEgen2}
\ee
We now proceed in turn to recast these two sets of BAEs in the form of
NLIEs.

\subsection{The first set of BAEs (\ref{BAEgen1}) and an auxiliary
inhomogeneous mixed spin chain}\label{subsec:first}

An important hint on how to analyze the first set of BAEs
(\ref{BAEgen1}) comes from rewriting it in the obviously equivalent
form
\be
\left(l_{j}+i\over l_{j}-i\right)^{L}
\prod_{k=1}^{M_{u}}
{l_{j}-u_{k}+i/2 \over l_{j}-u_{k}-i/2}=
\prod_{k=1 \atop k\ne j}^{M_{l}}
{l_{j}-l_{k}+i \over l_{j}-l_{k}-i} \,.
\ee
We recognize these as the BAEs for an inhomogeneous ``mixed'' spin
chain which has two types of spins: spin-1 and spin-1/2, with $L$ of
the former and $M_{u}$ of the latter.  (See, e.g., \cite{dVW}.)
Moreover, the latter have associated ``inhomogeneities'' $i u_{k}$,
$k=1, \ldots, M_{u}$.

We therefore consider an auxiliary integrable inhomogeneous mixed
quantum spin chain, where the number of spin-1 and spin-1/2
``quantum'' spaces are given respectively by $L$ and $M_u$; and with
spectral parameter inhomogeneities $i u_{k}$ only for the spin-1/2
spins.  This chain has two relevant transfer matrices $T_1(x)\,,
T_2(x)$, corresponding to ``auxiliary'' spaces which are spin-1/2
(2-dimensional) and spin-1 (3-dimensional), respectively.

We find by standard methods that the eigenvalues of these transfer
matrices (which we denote by the same notation) are given by
\footnote{Note that in place of the standard spectral parameter $u$,
we introduce $u=i x$.}
\be
T_1(x) &=& \psi(x-i/2) \phi(x-i ) \frac{Q(x+i )}{Q(x)}
+   \psi(x+i/2) \phi(x+i ) \frac{Q(x-i )}{Q(x)}\,, \label{t1expr} \\
T_2(x) &=&
\psi(x) \psi(x-i) \phi(x-i/2)  \phi(x-3i/2)  \frac{Q(x+ 3i/2)}{Q(x-i/2)}  \non \\
&+& \psi(x)^2 \phi(x-i/2)  \phi(x+i/2) \frac{Q(x+ 3i/2) Q(x-3i/2)}
 {Q(x+ i/2) Q(x-i/2)} \np
&+&  \psi(x) \psi(x+i)  \phi(x+i/2)  \phi(x+3i/2)
\frac{Q(x- 3i/2)}{Q(x+i/2)}
\non \\
&:=&\lambda_1(x)+\lambda_2(x)+\lambda_3(x) \,, \label{t2expr}
\ee
where
\be
\phi(x) =x^L \,,  \quad \psi(x)= \prod_{j=1}^{M_u}(x-u_j)   \,, \quad
Q(x) =\prod_{j=1}^{M_{l}} (x- l_j) \,.
\ee
Indeed, the BAEs obtained by demanding that $T_1(x)$ be analytic at
$x=l_{j}$ (zeros of $Q(x)$) coincide with (\ref{BAEgen1}).

Evidently $T_2(x)$ has the common factor $\psi(x)$, which has
``trivial'' zeros. We therefore  introduce the renormalized $T_2$,
$$
T_2(x) =\psi(x)\, T^{(r)}_2(x) \,.
$$
We note that
\be
S_{1}=S^{z}_{1}=L + \frac{M_{u}}{2} - M_{l} \,,
\ee
and we recall that the ``energy'' is given by (\ref{gammagen}).

\subsubsection{Physical degrees of freedom}

For simplicity, we restrict $u_k$ to be real, and $M_u=1, 2$.
For now, we also assume that $u_k$ are given by hand, with (in the case
$M_u=2$)  $u_1=-u_2$. We shall discuss how they should be determined
later in Sec. \ref{subsec:second}.

Numerical studies for small values of $L$ suggest that:
\begin{itemize}

\item For $M_u=1$, the lowest energy state in the $S^{z}_{1}=1/2$
sector is characterized by a single zero ($\vartheta_\alpha$) of
$T_1(x)$, and a single zero ($\theta_h$) of $T^{(r)}_2(x)$.  Both of
these zeros lie on the real axis.

\item For $M_u=2$, the lowest energy state is in the $S^{z}_{1}=0$ sector.
In the ``physical strip'' ($-1/2 \le \Im m\, x \le 1/2$), $T_1(x)$ and
$T^{(r)}_2(x)$ are free from zeros.

\item For $M_u=2$, the second-lowest energy state is in the $S^{z}_{1}=1$
sector.  It is characterized by two zeros ($\vartheta_\alpha$) of
$T_1(x)$ and two zeros ($\theta_h$) of $T^{(r)}_2(x)$.  These zeros
lie on the real axis.

\end{itemize}

These observations suggest that three sets of real parameters are
needed to describe the physical degrees of freedom: $u_j\,,
\vartheta_{\alpha}\,, \theta_h$.  The first and third parameters
correspond to magnon and spinon rapidities, respectively.  The second
parameter, which seems to correspond to excitation of the RSOS degree
of freedom, is not discussed in \cite{BFHZ}.

\subsubsection{The auxiliary functions and algebraic relations among them}

As in previous studies \cite{Su1, Su2}, we introduce a pair of
auxiliary functions
\be
\mathfrak{b}_1(x) := \frac{\lambda_1(x)+\lambda_2(x)}{\lambda_3(x)}
\qquad \Im m\, x \ge 0 \,,\qquad
\bar{\mathfrak{b}}_1(x) :=
\frac{\lambda_2(x)+\lambda_{3}(x)} {\lambda_{1}(x)}  \qquad  \Im m\,
x \le 0 \,,
\label{defb}
\ee
where $\lambda_{i}(x)$ are defined in (\ref{t2expr}).
They are free from zeros and poles near the real axis.
This will be apparent from the following representations,
\be
\mathfrak{b}_1(x) &=&
\frac{\phi_{-1/2}}{\psi_1 \phi_{3/2} \phi_{1/2}}
\frac{Q(x+3i/2)}{Q(x-3i/2)} T_1(x-i/2) \,, \non \\
\bar{\mathfrak{b}}_1(x) &=&
\frac{\phi_{1/2}}{\psi_{-1} \phi_{-3/2} \phi_{-1/2}}
\frac{Q(x-3i/2)}{Q(x+3i/2)} T_1(x+i/2) \,.
\label{bT1rel}
\ee
We have introduced here the abbreviated notation $\phi_a:= \phi(x+
ia)$, and similarly for $\psi$, which we shall use throughout this part
of the paper.

At this stage, there seems to be no reason why the two auxiliary
functions should be introduced in the corresponding half planes.  This
will become clear at a later stage.

The upper-case functions are also introduced:
$\mathfrak{B}_1(x)= 1+\mathfrak{b}_1(x)$,
$\bar{\mathfrak{B}}_1(x)= 1+\bar{\mathfrak{b}}_1(x)$, and the
following relations are also useful:
\be
T^{(r)}_2(x) &=& \psi_{1} \phi_{1/2} \phi_{3/2}
\frac{Q(x-3i/2)}{Q(x+i/2)}  \mathfrak{B}_1(x)   \label{T2B} \\
&=& \psi_{-1}  \phi_{-1/2} \phi_{-3/2}
\frac{Q(x+3i/2)}{Q(x-i/2)}
\bar{\mathfrak{B}}_1(x)  \,. \label{T2BB}
\ee
Apparently $\mathfrak{B}_1(x)$ vanishes at $x= \theta_h$,  but it remains nonzero
at $x=u_j$.

We now define the most important functions,
\begin{align}
\mathfrak{b}(x) &= \mathfrak{b}_1(x+i\epsilon)\,, &
\mathfrak{B}(x) &= \mathfrak{B}_1(x+i\epsilon)   &\Im m\, x \ge 0 \,, \\
\bar{\mathfrak{b}}(x) &= \bar{\mathfrak{b}}_1(x-i\epsilon)\,, &
\bar{\mathfrak{B}}(x) &= \bar{\mathfrak{B}}_1(x-i\epsilon)    &\Im
m\, x \le 0 \,.
\end{align}
Here $\epsilon$ denotes a positive quantity which is slightly larger
than the deviation of the 2-strings from their ``perfect'' positions.
Therefore $\mathfrak{B}$ would possess zeros (due to the factor
$Q(x+i/2)$ in (\ref{T2B})) slightly below the real axis if it were
defined in the whole complex plane.  The function $\mathfrak{B}$ is,
however, defined only in the upper half plane (including the real
axis).

Another auxiliary function originates from the so-called fusion
formula that relates the two transfer matrices,
\be
T_1(x-i/2)\, T_1(x+i/2)= \psi_1 \psi_{-1} \phi_{3/2} \phi_{-3/2} +
\psi_0 T^{(r)}_2(x) \,,
\ee
which can be verified using (\ref{t1expr}) and (\ref{t2expr}).
For later convenience, we renormalize $T_1(x)=
\prod_{\alpha=1}^{N_{\vartheta}} \tanh
\frac{\pi}{2} (x-\vartheta_{\alpha})\, T_1^{(r)}(x)$, and
rewrite the above in the form
\be
T_1^{(r)} (x-i/2)\, T_1^{(r)}(x+i/2) &=&
\psi_1 \psi_{-1} \phi_{3/2} \phi_{-3/2} + \psi_0 T^{(r)}_2(x) \non \\
&=& \psi_1 \psi_{-1} \phi_{3/2} \phi_{-3/2}\, Y(x) \,,
\label{T1T1}
\ee
where we have defined the auxiliary functions
\be
y(x) := \frac{\psi_0}{\psi_1 \psi_{-1} \phi_{3/2} \phi_{-3/2}}
T^{(r)}_2(x) \,, \qquad
Y(x) := 1 + y(x) \,.
\label{yfunc}
\ee
Since $y$ possesses zeros on the real axis due to $u_j$ and
$\theta_h$, we also define a renormalized function $y^{(r)}$
\be
y(x) = \prod_{j=1}^{M_{u}} \tanh \frac{\pi}{2}(x-u_j)
\prod_{h=1}^{N_{h}} \tanh
\frac{\pi}{2}(x-\theta_{h})\,  y^{(r)}(x) \,,
\ee
which obeys the functional relation
\be
y^{(r)}(x-i/2)\, y^{(r)}(x+i/2) = \mathfrak{B}_1(x+i/2)\,
\bar{\mathfrak{B}}_1(x-i/2) \,,
\label{yyrel}
\ee
as follows from (\ref{T2B}) and  (\ref{T2BB}).

\subsubsection{Derivation of NLIE}

The derivation of the NLIE can be most easily done in Fourier space.
For a smooth function  $f(x)$, we define
\be
\hat{f}[k]= \frac{1}{2\pi} \int_{-\infty}^{\infty} {\rm e}^{ikx}
f(x)\, dx \,,
\qquad
f(x) = \int_{-\infty}^{\infty} {\rm e}^{-ikx} \hat{f}[k]\, dk.
\ee
We also introduce the special notation
\be
\widehat{dl f}[k]= \frac{1}{2\pi}  \int_{-\infty}^{\infty} {\rm
e}^{ikx}  \left[ \ln f(x) \right]' dx
\label{specialnot}
\ee
which will be frequently used below.

It is convenient to introduce ``shifted'' $Q$ functions,
\be
q_1(x) := Q(x-i/2-i\epsilon) \,, \qquad  q_2(x) := Q(x+i/2+i\epsilon)
\,.
\ee
By definition, $q_1$ is Analytic and NonZero (ANZ) for $\Im m\,  x \le
0$, while
$q_2$ is ANZ for $\Im m\,  x \ge 0$.
We therefore have by Cauchy's theorem the important property
\be
\widehat{dl q}_2 [k > 0 ]  = \widehat{dl q}_1 [k < 0 ]  =0 \,.
\ee
Similarly,
\be
\widehat{dl \psi}_a [k > 0 ] = \widehat{dl \phi}_a [k > 0 ] =0
\mbox{  for  } a>0 \,, \quad
\widehat{dl \psi}_a [k < 0 ] = \widehat{dl \phi}_a [k < 0 ] =0
\mbox{  for  } a<0 \,.
\ee

\begin{figure}[htb]
    \centering
    \includegraphics[height=5cm]{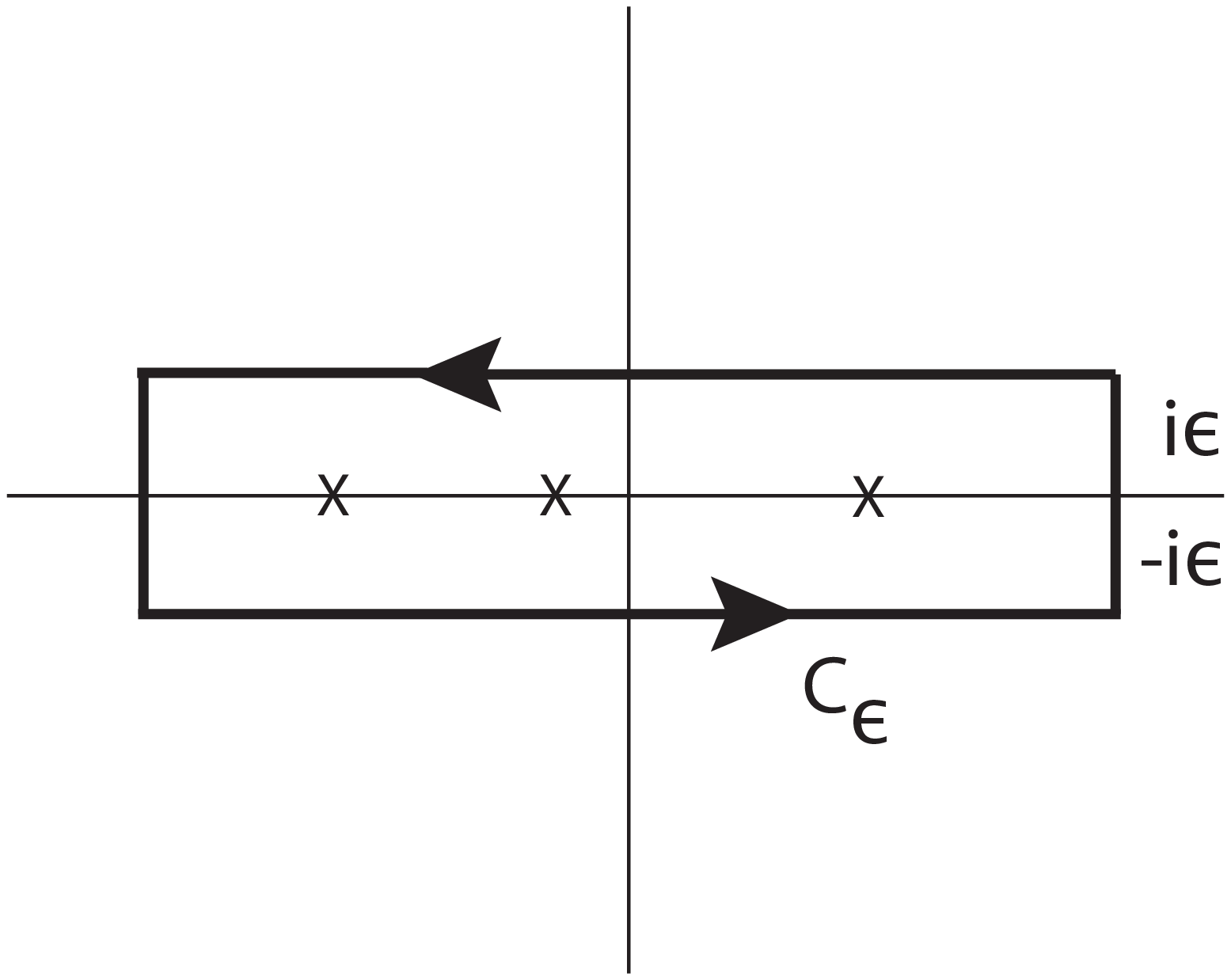}
    \caption[xxx]{\parbox[t]{0.25\textwidth}{
Integration contour}
    }
    \label{fig:contour}
\end{figure}

We slightly shift the arguments in  (\ref{T2B}), (\ref{T2BB})
\be
T^{(r)}_2(x+i \epsilon) &=& \psi_{1+\epsilon} \phi_{1/2+\epsilon} \phi_{3/2+\epsilon}
\frac{q_1(x-i+2i \epsilon)}{q_2(x)}  \mathfrak{B}(x)  \,, \label{T2B2} \\
T^{(r)}_2(x-i \epsilon) &=& \psi_{-1-\epsilon}  \phi_{-1/2-\epsilon} \phi_{-3/2-\epsilon}
\frac{q_2(x+i-2i \epsilon) }{q_1(x)}  \bar{\mathfrak{B}} (x) \,. \label{T2BB2}
\ee
We then use the result
\be
\frac{1}{2\pi}  \int_{C_{\epsilon}} {\rm e}^{i k x} \left[\ln
T_2^{(r)}(x)\right]' dx =i \sum_h {\rm e}^{ik\theta_h} \,,
\ee
where we choose the contour $C_{\epsilon}$ as in Figure
\ref{fig:contour},
and we obtain the following
\be
\widehat{dl q}_1[k>0] &=&
\frac{ \widehat{dl \psi}_{-1-\epsilon}+ \widehat{dl \phi}_{-3/2-\epsilon}
+\widehat{dl \phi}_{-1/2-\epsilon}}{1+{\rm e}^{-k}}
+\frac{\widehat{dl \bar{\mathfrak{B}}}[k]- {\rm e}^{-2k \epsilon} \widehat{dl \mathfrak{B}}[k]}
{1+{\rm e}^{-k}} \non \\
& & -i \sum_h \frac{ {\rm e}^{ik \theta_h -k \epsilon}}{1+{\rm e}^{-k}} \,, \label{q1sol} \\
\widehat{dl q}_2[k<0] &=&
\frac{ \widehat{dl \psi}_{1+\epsilon}+ \widehat{dl \phi}_{3/2+\epsilon}
+\widehat{dl \phi}_{1/2+\epsilon}}{1+{\rm e}^{k}}
+\frac{\widehat{dl \mathfrak{B}}[k]- {\rm e}^{2k \epsilon} \widehat{dl \bar{\mathfrak{B}}}[k]}
{1+{\rm e}^{k}} \non \\
& & + i \sum_h \frac{ {\rm e}^{ik \theta_h + k \epsilon}}{1+{\rm e}^{k}}
\,. \label{q2sol}
\ee

In addition, from (\ref{T1T1}), one  derives
\be
\widehat{dl T}_1[k] =\frac{ \widehat{dl \psi}_{\mp 1}
+\widehat{dl \phi}_{\mp 3/2}}{ {\rm e}^{k/2}+ {\rm e}^{-k/2}  }
+ \frac{ \widehat{dl Y}  }{ {\rm e}^{k/2}+ {\rm e}^{-k/2}  }
+i\sum_{\alpha}  \frac{ {\rm e}^{ik \vartheta_{\alpha} + k/2} }{ {\rm
e}^{k/2}+ {\rm e}^{-k/2} } \,,
\label{T1FT}
\ee
where $- (k>0)$ and $+ (k<0)$.

We shift the arguments in (\ref{bT1rel})
\be
\mathfrak{b} (x) &=&
\frac{\phi_{-1/2+\epsilon }}{\psi_{1+\epsilon}  \phi_{3/2+\epsilon} \phi_{1/2+\epsilon}}
\frac{q_2(x+i)}{q_1(x-i+2 i \epsilon )} T_1(x-i/2+i\epsilon) \,, \non \\
\bar{\mathfrak{b}} (x) &=&
\frac{\phi_{1/2-\epsilon }}{\psi_{-1-\epsilon} \phi_{-3/2-\epsilon} \phi_{-1/2-\epsilon}}
\frac{q_1(x-i)}{q_2(x+i-2i \epsilon)} T_1(x+i/2-i\epsilon)\,,
\label{bT1rel2}
\ee
and then take the Fourier transformation.  The substitution of
(\ref{q1sol} ), (\ref{q2sol} ) and (\ref{T1FT}) into the resultant
transformation then leads to the NLIE in Fourier space,
\be
\widehat{dl b}[k>0] &=&
\frac{\widehat{dl \phi}_{-1/2+\epsilon} }{1+{\rm e}^{-k}} +
i \sum_h \frac{ {\rm e}^{ik\theta_h + \epsilon k} }{1+ {\rm e}^{k} } +
i  \sum_\alpha \frac{ {\rm e}^{ik\vartheta_{\alpha}+ \epsilon k} }
{ {\rm e}^{k/2}+ {\rm e}^{-k/2} }  \non \\
&+& \frac{  {\rm e}^{-k/2+\epsilon k}}{ {\rm e}^{k/2}+ {\rm e}^{-k/2} } \widehat{dl Y}[k]
+  \frac{ 1} { {\rm e}^{k}+ 1} ( \widehat{dl \mathfrak{B}} [k]
-{\rm e}^{2k \epsilon} \widehat{dl \bar{\mathfrak{B}}} [k] ) \,,  \\
\widehat{dl b}[k<0] &=&
-\frac{\widehat{dl \phi}_{1/2+\epsilon} }{1+{\rm e}^{k}} +
i \sum_h \frac{ {\rm e}^{ik\theta_h + \epsilon k} }{1+ {\rm e}^{-k} } +
i \sum_\alpha \frac{ {\rm e}^{ik\vartheta_{\alpha}+ \epsilon k} }{
{\rm e}^{k/2}+ {\rm e}^{-k/2} }  \non \\
&+& \frac{  {\rm e}^{-k/2+\epsilon k}}{ {\rm e}^{k/2}+ {\rm e}^{-k/2} } \widehat{dl Y}[k]
+ \frac{ 1} { {\rm e}^{-k}+ 1 } ( \widehat{dl \mathfrak{B}} [k]
- {\rm e}^{2 k \epsilon}\widehat{dl \bar{\mathfrak{B}}} [k] ) \,.
\ee
Interestingly, although a contribution from the inhomogeneities ($\psi$) appeared
during the calculation, it cancelled in the final form.
An equation for $y$ is immediately derived from (\ref{yyrel}),
\be
\widehat{dl y}[k] &=& i \sum_h \frac{ {\rm e}^{ik\theta_h} }{1+ {\rm e}^{-k} }
+ i \sum_j \frac{ {\rm e}^{ik u_{j}} }{1+ {\rm e}^{-k} } \non \\
&+&  \frac{{\rm e}^{k(1/2-\epsilon)}} {{\rm e}^{k/2} + {\rm e}^{-k/2}}
\widehat{dl \mathfrak{B}}[k]
+ \frac{{\rm e}^{-k(1/2-\epsilon)}} {{\rm e}^{k/2} + {\rm e}^{-k/2}}
\widehat{dl \bar{\mathfrak{B}}}[k] \,. \label{yk}
\ee

In the original coordinate space, the resultant equations read
\be
\ln \mathfrak{b}(x)  &=& i D_b (x+i\epsilon)
+\int_{-\infty}^{\infty} G_s(x-x') \ln \mathfrak{B}(x')\, dx'
-\int_{-\infty}^{\infty} G_s(x-x'+2 i \epsilon) \ln
\bar{\mathfrak{B}}(x')\, dx'   \non \\
&+& \int_{-\infty}^{\infty} K(x-x'-i/2 +i\epsilon) \ln Y
(x')\, dx'    \,, \label{bNLIE2}  \\
\ln y (x)  &=&  i D_y (x)
+ \int_{-\infty}^{\infty} K(x-x'+i/2 -i\epsilon )
\ln \mathfrak{B}(x')\,  dx'   \non  \\
&+& \int_{-\infty}^{\infty} K(x-x'-i/2 +i\epsilon )
\ln \bar{\mathfrak{B}}(x')\, dx'  \,,  \label{y1NLIE2}
\ee
where
\be
G_s(x) = \frac{1}{2\pi} \int_{-\infty}^{\infty}
\frac{  {\rm e}^{-ikx} }{ 1+ {\rm e}^{|k|}} dk  \,, \qquad
K(x)  = \frac{1}{2\pi}  \int_{-\infty}^{\infty}
\frac{1 }{ 2\cosh (k/2)}   {\rm e}^{-ikx}\,  dk
=\frac{1}{2\cosh(\pi x)} \,.
\ee
The source term in (\ref{bNLIE2}) consists of the bulk (``driving'') contribution
and the contribution from the hole excitations,
\be
D_b (x) =D^{(b)}_{\rm bulk}(x) + D^{(b)}_{\rm hole}(x)
-\frac{\pi}{2}N_{h} \,,
\label{Db1}
\ee
where
\be
D^{(b)}_{\rm bulk}(x)  =  L \chi_K(x) \,, \qquad
D^{(b)}_{\rm hole}(x)  =
\sum_{\alpha=1}^{N_{\vartheta}}  \chi_K(x-\vartheta_{\alpha})
+\sum_{h=1}^{N_{h}}  \chi(x-\theta_h) \,,
\label{Db2}
\ee
and
\be
\chi'_K(x) = 2\pi K(x) \,, \qquad \chi'(x)= 2\pi G_s(x) \,.
\label{chiKchi}
\ee
In particular, on suitable domains (containing the positive real
axis),
\be
\chi_K(x) = \frac{1}{i}\ln \tanh(\pi(x-i/2)/2)  =
i\ln \frac{\sinh(\pi(x+i/2)/2)}{\sinh(\pi(x-i/2)/2)}
+ \frac{\pi}{2}
= \arctan(\sinh(\pi x)) -\frac{\pi}{2} \,, \non \\
\ee
where $\chi_K(0) \equiv -\pi/2$, and also $\chi(0) \equiv 0$.
The source term in (\ref{y1NLIE2}) is given by
\be
D_y(x)
&=&  \sum_{h=1}^{N_{h}} \chi_K(x-\theta_h +i/2)
+ \sum_{j=1}^{M_{u}} \chi_K(x-u_j +i/2) \,.
\label{Dy}
\ee

The parameters $(u_j, \vartheta_{\alpha}, \theta_h)$ must actually be
determined again by NLIEs.  Indeed, (\ref{T2B}) implies that the hole
rapidities $\theta_h$ are determined by
\be
\mathfrak{b}(\theta_h-i\epsilon)=\mathfrak{b}_1(\theta_h) = -1 \,,
\label{thetah}
\ee
which also leads to the determination of the spinon-spinon and
spinon-magnon scattering matrices, as discussed in Sec.
\ref{sec:Smatrix}.

In order to fix the parameters $\vartheta_{\alpha}$, we need another
NLIE. We consider the most natural auxiliary function ${\mathfrak a}(x)$,
defined by \footnote{As discussed further in Sec.
\ref{subsec:counting}, one can verify numerically that $\frac{1}{i}\ln
{\mathfrak a}(x)$ and also $\Re\, e \left[\frac{1}{i}\ln
{\mathfrak b}(x) \right]$ are increasing functions of
$x$.}
\be
{\mathfrak a}(x) := \frac{\lambda_1(x+i/2)} {\lambda_2(x+i/2)}
 =\frac{\lambda_2(x-i/2)} {\lambda_3(x-i/2)} \,,
\label{auxfunca}
\ee
where again $\lambda_{i}(x)$ are defined in (\ref{t2expr}). From
(\ref{t1expr}) we have
\be
T_{1}(x) = \psi_{1/2} \phi_{1} \frac{Q(x-i)}{Q(x)}\left[1 +
{\mathfrak a}(x) \right] \,.
\ee
Hence, the zeros of $T_1$ on the real axis $\vartheta_{\alpha}$ satisfy
\be
\mathfrak{a}(\vartheta_{\alpha})=-1 \,.
\label{varthetah}
\ee

We omit the derivation of the NLIE for $\mathfrak{a}(x)$ for $|\Im
m\,  x| < 1/2$, which is similar to the one for the trigonometric and
homogeneous case considered in \cite{Su2}. The result is
\be
\ln \mathfrak{a}(x) = i D_a(x)
+ \int_{-\infty}^{\infty} K(x-x'-i\epsilon) \ln \mathfrak{B}(x')\, dx'
- \int_{-\infty}^{\infty} K(x-x'+i\epsilon) \ln \bar{\mathfrak{B}}(x')\, dx'
\,, \label{aNLIE1}
\ee
where the source term is given by
\be
D_a(x) =  \sum_{h=1}^{N_{h}} \chi_K(x-\theta_h )
+\sum_{j=1}^{M_{u}} \chi_K(x-u_j ) \,.
\label{Da}
\ee

\subsection{The second set of BAEs (\ref{BAEgen2})}\label{subsec:second}

We finally consider an equation to fix the magnon rapidities $u_j$.
For this purpose, we propose an expression for the transfer matrix
eigenvalues similar to the one for the $su(3)$ spin chain,
\footnote{We expect that, starting from a suitable $su(2,1)$ $R$
matrix, a transfer matrix can be constructed with eigenvalues
(\ref{tau}).  However, we have not attempted to carry out this construction.}
\be
\tau(x) &=&
\phi(x-i) \frac{Q(x+i)}{Q(x)} +
\phi(x+i)  \frac{\psi(x+i/2)}{\psi(x-i/2)} \frac{Q(x-i)}{Q(x)}
+\phi(x-2i)\frac{\psi(x-3i/2)}{\psi(x-i/2)} \non \\
&:=&\tau_1(x)+\tau_2(x)+\tau_3(x) \,. \label{tau}
\ee
Indeed, demanding analyticity of $\tau(x)$ at
$x=l_{j}$ (zeros of $Q(x)$) gives the BAEs (\ref{BAEgen1}), while
demanding analyticity at $x=u_{j}+i/2$  (zeros of $\psi(x-i/2)$)
gives the BAEs (\ref{BAEgen2}).

Because of its similarity to the $su(3)$ transfer matrix eigenvalue,
we shall assume that $\tau(x)$ is ANZ in the strip $-1/2 \le \Im m\, x \le
1/2$, which is indeed the analyticity property for the $su(3)$ case.  This
assumption can in principle be checked numerically for small values of
$L$.  However, we have so far not succeeded to do so, due to the
difficulty of finding numerical solutions of the BAEs (\ref{BAEgen1}),
(\ref{BAEgen2}).

This assumption leads to a simple determination of $u_j$ as follows.
Let us consider an auxiliary function introduced in studies of the
supersymmetric $tJ$ model \cite{JKS} and the $su(3)$ vertex model
\cite{FK},
\be
\mathfrak{c}(x) := \frac{\tau_3(x+i/2)}{\tau_1(x+i/2) + \tau_2(x+i/2)}
\label{cauxfunc}
\,.
\ee
It is easy to check that this can be rewritten in terms of $T_1(x)$ in
(\ref{t1expr}),
\be
\mathfrak{c}(x) = \frac{\psi(x-i) \phi(x-3i/2)}{T_1(x+i/2)} \,.
\ee
We then have
\be
\mathfrak{c}(u_j) = -1 \,,
\label{uj}
\ee
which follows from
\be
\mathfrak{C}(x)= 1+\mathfrak{c}(x)=
\frac{\tau(x+i/2)}{\tau_1(x+i/2)+\tau_2(x+i/2)}
=\frac{\tau(x+i/2)\, \psi(x)}{T_1(x+i/2)} \,.
\ee
From our above assumption on the analyticity of $\tau(x)$, the
zeros of $\mathfrak{C}(x)$ near the real axis are determined by those
of $\psi(x)$, namely $u_j$.

The NLIE for $\mathfrak{c}$ is obtained from the knowledge of $T_1$.
The result is
\be
\ln\mathfrak{c}(x) =  i D_c(x) - \int_{-\infty}^{\infty} K(x-x'+i/2)
\ln Y(x')\, dx'  \,, \label{lnc}
\ee
where the source term is given by
\be
D_c(x)
&=& L  \chi_2(x) +  \sum_{j=1}^{M_{u}} \chi_{3/2} (x-u_j)
+  \sum_{\alpha=1}^{N_{\vartheta}} \chi_K(x-\vartheta_{\alpha})
- \frac{\pi}{2}(L+M_{u})\,,
\label{Dc}
\ee
and
\be
\chi_{a}'(x) = 2\pi K_{a}(x)\,,  \quad
K_{a} (x) =\frac{1}{2\pi} \int_{-\infty}^{\infty}
\frac{{\rm e}^{-a|k|-ikx}}{ 2\cosh \frac{k}{2} }dk\,,  \quad
\chi_{a}(0)\equiv 0\,,
\quad a=3/2\,, 2 \,.
\label{kernelKa}
\ee

\subsection{Counting functions and counting
equations}\label{subsec:counting}

So-called counting equations relating the various types of Bethe roots
and excitations in a given state can be derived from corresponding
counting functions associated with the auxiliary functions .  These
counting equations help determine the spins of the excitations.

We continue to restrict to the case of real $u$-roots and no
$r$-roots.  As in previous studies \cite{NLIE2, Su2, HRS}, it is
convenient to classify $l$-roots according to their imaginary parts as
follows:
\begin{description}
  \item[2-strings]: pairs of
complex-conjugate roots $x_{j} \pm i y_{j}$ with $0 < y_{j}-1/2 <\!\!<
1$, $j=1, \ldots, M_{2}/2$
  \item[real roots]: $\Im m\, l_{j}=0$, $j=1, \ldots, M_{real}$
  \item[inner roots]: $|\Im m\, l_{j}| < 1/2$, $j=1, \ldots, M_{I}$
  \item[close roots]: $1/2 < |\Im m\, l_{j}| < 3/2$, $j=1, \ldots,
  M_{C}$
  \item[wide roots]: $|\Im m\, l_{j}| > 3/2$, $j=1, \ldots,
   M_{W}$
\end{description}
Hence,
\be
M_{l} = M_{real} + M_{2} + M_{I} + M_{C} + M_{W} \,.
\ee
It is also convenient to introduce the functions
\be
\theta_{\mp}(x,\alpha) = \frac{1}{i} \ln\left(\mp
\frac{x-i\alpha}{x+i\alpha}\right)\,.
\ee
Note that $\theta_{-}(x,\alpha) = 2 \arctan (x/\alpha)$ has branch
points in the complex $x$ plane at $x=\pm i \alpha$; following
\cite{NLIE2}, we choose the corresponding branch cuts to be parallel
to the real axis, extending from $i\alpha$ to $+\infty + i\alpha$, and
from $-\infty - i\alpha$ to $-i\alpha$. This function has a
discontinuity of $-2\pi$ when crossing the cuts from below. Similarly,
we add to $\theta_{+}(x,\alpha)$ a $2\pi$-discontinuity at
$x=0$ so that it is a continuous function of $x$.

We define the counting function $z_{\mathfrak{a}}(x)$ associated with
the auxiliary function $\mathfrak{a}(x)$ (\ref{auxfunca}) by
\be
z_{\mathfrak{a}}(x)=\frac{1}{i}{\rm Log}\, \mathfrak{a}(x) = L
\theta_{-}(x,1) -\sum_{j=1}^{M_{l}}\theta_{-}(x-l_{j},1)
+\sum_{j=1}^{M_{u}}\theta_{+}(x-u_{j},1/2)\,.
\label{acountingfunc}
\ee
We have verified numerically for various states that
$z_{\mathfrak{a}}(x)$ is a continuous increasing function of $x$.
This function ``counts'' zeros of $T_{1}(x)$ and real $l$-roots.  That
is,
\be
z_{\mathfrak{a}}(x_{j}) = 2\pi I_{j}^{\mathfrak{a}} \,,
\label{acountingfunc2}
\ee
where $I_{j}^{\mathfrak{a}}$ is integer ($S_{1}-S_{2}$ odd) or half-odd
integer ($S_{1}-S_{2}$ even) if $x_{j}$ is a zero of $T_{1}(x)$ or a real $l$-root.
Defining integers or half-odd integers $I_{max}^{\mathfrak{a}}$ and
$I_{min}^{\mathfrak{a}}$ by
\be
z_{\mathfrak{a}}(+\infty) &=& 2\pi \left(I_{max}^{\mathfrak{a}} +
\frac{1}{2}\right) \,, \non \\
z_{\mathfrak{a}}(-\infty) &=& 2\pi \left(I_{min}^{\mathfrak{a}} -
\frac{1}{2}\right) \,,
\ee
it follows from (\ref{acountingfunc}) and
(\ref{acountingfunc2}), respectively, that
\be
I_{max}^{\mathfrak{a}} - I_{min}^{\mathfrak{a}} +1 &=& S_{1} + S_{2} +
M_{b}   \non \\
&=& N_{\vartheta} +  M_{real} \,,
\ee
where $M_{b}$ is the number of $l$-roots $l_{j}$ with $|\Im m\, l_{j}| >
1$. We therefore
arrive at the first counting equation
\be
N_{\vartheta} = S_{1} + S_{2} + M_{b} -  M_{real} \,.
\label{counting1}
\ee

Similarly, we define the counting function $z_{\mathfrak{b}}(x)$ associated with
the auxiliary function $\mathfrak{b}_{1}(x)$ (\ref{defb}) by
\be
z_{\mathfrak{b}}(x) &=& \Re e\, \frac{1}{i}{\rm Log}\,
\mathfrak{b}_{1}(x)  = \Re e\, \Big\{ \frac{1}{i}\ln \left[1 +
\frac{1}{\mathfrak{a}(x-i/2)} \right]
+\sum_{j=1}^{M_{u}}\theta_{+}(x-u_{j},1) \non \\
&+&L\left[\theta_{-}(x,1/2)+\theta_{-}(x,3/2)\right]
-\sum_{j=1}^{M_{l}}\left[\theta_{-}(x-l_{j},1/2)+
\theta_{-}(x-l_{j},3/2)\right] \Big\}
\,.
\label{bcountingfunc}
\ee
The presence of the first term generally requires the introduction of
further discontinuities.  We have verified numerically that
$z_{\mathfrak{b}}(x)$ is also a continuous increasing function of $x$.
This function ``counts'' zeros of $T_{2}^{(r)}(x)$ and centers of
2-strings and inner pairs.  Proceeding as before, we find
\be
I_{max}^{\mathfrak{b}} - I_{min}^{\mathfrak{b}} +1 &=& 2S_{1} +
M_{C} + 2 M_{W} + (M_{2} + M_{I})/2 \non \\
&=& N_{h} + (M_{2} + M_{I})/2 \,.
\ee
We therefore arrive at the second counting equation
\be
N_{h} = 2S_{1} + M_{C} + 2M_{W} \,.
\label{counting2}
\ee

Finally, we define the counting function $z_{\mathfrak{c}}(x)$ associated with
the auxiliary function $\mathfrak{c}(x)$ (\ref{cauxfunc}) by
\be
z_{\mathfrak{c}}(x) &=& \Re e\, \frac{1}{i}{\rm Log}\,
\mathfrak{c}(x)  = \Re e\, \Big\{ -\frac{1}{i}\ln \left[1 +
\mathfrak{a}(x+i/2) \right]
+\sum_{j=1}^{M_{u}}\theta_{+}(x-u_{j},1) \non \\
&+&L \theta_{-}(x,3/2)
-\sum_{j=1}^{M_{l}} \theta_{-}(x-l_{j},1/2) \Big\}
\,.
\label{ccountingfunc}
\ee
We have verified numerically (using for the first term the same
discontinuities introduced for the first term in
(\ref{bcountingfunc})) that $z_{\mathfrak{c}}(x)$ is a continuous
increasing function of $x$.  Assuming
\be
I_{max}^{\mathfrak{c}} - I_{min}^{\mathfrak{c}} +1 = L + M_{u} -
M_{real} - (M_{2} + M_{I})/2 \,,
\ee
which can also be verified numerically, we recover the result
\be
M_{u} = 2S_{2} \,.
\label{counting3}
\ee

\section{Spin, energy and momentum of excitations}\label{sec:energymom}

We now compute the excitations' spin, energy and momentum, which enter into
the computation of the $S$ matrix.  Our results agree (except for some
minor discrepancies) with those obtained previously using the string
hypothesis.

We can infer the spins of the excitations with the help of the
counting equations found in Sec. \ref{subsec:counting}.
The second counting equation (\ref{counting2}) implies that a spinon
has $S_{1}=1/2$. Indeed, $N_{h}=1$ requires $S_{1}=1/2$ (and
$M_{C}=M_{W}=0$); $N_{h}=2$ requires either $S_{1}=0$ or $S_{1}=1$,
etc. Note that all the terms on the RHS of (\ref{counting2})
are nonnegative. Evidently, a spinon also has $S_{2}=0$.
The fact that a spinon has spin-1/2 was found
using the string hypothesis by Takhtajan \cite{Ta}.

Similarly, the third counting equation (\ref{counting3}) implies that
a magnon has $S_{2}=1/2$, and evidently $S_{1}=0$.  This result was
found using the string hypothesis by Beisert {\it et al.} \cite{BFHZ}.

The spin of the $\vartheta$ particle is not determined by the first
counting equation  (\ref{counting2}), since not all the terms on the RHS
are nonnegative. Nevertheless, an analysis of various examples
suggests that this particle has $S_{1}=S_{2}=0$.

By the definition in \cite{BFHZ}, the energy ($E$) is related to the
anomalous dimension (\ref{gammagen}) by $\gamma = {\alpha_{s}
N_{c}\over 2\pi} E$, and is therefore given by \footnote{For
convenience, we drop the constant term $7L/6$ in the expression for $E$.
This definition of energy is (for the $l$-roots) a factor 2 larger than
the one in \cite{Ta}.}
\be
E = - \sum_{j=1}^{M_{l}}{2\over l_{j}^{2}+1}
+ \sum_{j=1}^{M_{u}}{3\over u_{j}^{2}+9/4} \,.
\label{energy}
\ee
We can relate this to the derivate of the eigenvalue $T_{2}(x)$
(\ref{t2expr}) at $x=i/2$,
\be
E = i {d\over dx} \ln T_{2}(x) \Big\vert_{x=i/2} - {3L\over 2} +
\sum_{j=1}^{M_{u}}\left[{3\over u_{j}^{2} + 9/4}
+i\left({1\over u_{j} -i/2}+{1\over u_{j}-3i/2}\right)  \right] \,.
\ee
Recalling the definition of the auxiliary function $y(x)$ (\ref{yfunc}),
we see that
\be
E = i {d\over dx} \ln y(x) \Big\vert_{x=i/2} - 2L +
\sum_{j=1}^{M_{u}}\left({3\over u_{j}^{2} + 9/4}
-{1\over u_{j}^{2} + 1/4}
 \right) \,.
\ee
We observe from (\ref{specialnot}) that
\be
{d\over dx} \ln y(x) = \int_{-\infty}^{\infty}dk\ e^{-i k x}\widehat{dl y}[k] \,,
\ee
and substitute our result for $\widehat{dl y}[k]$ (\ref{yk}) to obtain
\be
E= -2L + \sum_{j=1}^{M_{u}} \left[ {\pi\over \cosh ( \pi u_{j})}
+ {3\over u_{j}^{2} + 9/4} -{1\over u_{j}^{2} + 1/4}
\right] + \sum_{h} {\pi\over \cosh ( \pi \theta_{h})} + \ldots \,,
\label{energyresult}
\ee
where the ellipsis (\ldots) represents the Casimir energy contribution.
We conclude that the energy of a spinon is
\be
\varepsilon_{h}(\theta) = {\pi\over \cosh ( \pi \theta)} \,,
\label{spinonenergy}
\ee
and the energy of a magnon is
\be
\varepsilon_{u}(u) = {\pi\over \cosh ( \pi u)}
+ {3\over u^{2} + 9/4} -{1\over u^{2} + 1/4} \,,
\label{magnonenergy}
\ee
in agreement with Eqs.  (6.15), (6.32) in Beisert {\it et al.}
\cite{BFHZ}, respectively, up to a factor 2.  The spinon result
(\ref{spinonenergy}) was first found by Takhtajan \cite{Ta}.
We remark that
\be
\varepsilon_{u}(u) = 2\pi K_{2}(u) \,,
\label{magnonenergy2}
\ee
where $K_{2}(u)$ is the kernel introduced in (\ref{kernelKa}).
Evidently there is no $\vartheta$-dependent contribution in
(\ref{energyresult}), which implies that the $\vartheta$ ``particle''
does not carry energy.

The momentum is given by \footnote{This definition of momentum differs
(for the $l$-roots) by an overall sign from the one in \cite{Ta}.}
\be
P = {1\over i}\left[ \sum_{j=1}^{M_{l}} \ln \left( {l_{j} + i\over
l_{j} - i} \right)
+ \sum_{j=1}^{M_{u}} \ln \left(
\frac{u_{j} - 3i/2}{u_{j} + 3i/2} \right) \right] \quad (\mbox{mod  } 2\pi) \,.
\ee
We can evaluate it in similar fashion. Indeed, we find that
\be
P = {1\over i} \ln y(i/2)
+ {1\over i}\sum_{j=1}^{M_{u}} \left[ \ln e_{-3}(u_{j}) + \ln
e_{1}(u_{j}) \right] +  L \pi \,,
\ee
where we have introduced the notation
\be
e_{n}(u) = \frac{u + i n/2}{u - i n/2} \,.
\ee
Proceeding as before, we arrive at the result
\be
P= L \pi + \sum_{j=1}^{M_{u}} \left[ \chi_{K}(u_{j}) +
q_{3}(u_{j}) - q_{1}(u_{j})
\right] + \sum_{h} \chi_{K}(\theta_{h}) + \ldots \,,
\ee
where $\chi_{K}(x)$ is defined in (\ref{chiKchi}),
and $q_{n}(x)$ is defined by
\be
q_{n}(x) = \pi + i \ln e_{n}(x) \quad n> 0 \,, \qquad
q_{-n}(x) = -q_{n}(x) \,,
\qquad  q_{0}(x) = 0
\,. \label{qn}
\ee
It is an odd function of $x$, and satisfies
\be
q_{n}(x) = 2 \arctan \left(2x/n\right) \quad n \ne 0 \,.
\ee
We conclude that the momentum of a spinon is
\be
p_{h}(\theta) = \chi_{K}(\theta) \,,
\label{spinonmomentum}
\ee
and the momentum of a magnon is
\be
p_{u}(u) = \chi_{K}(u)  + q_{3}(u) - q_{1}(u)  \,,
\label{magnonmomentum}
\ee
in agreement with Eqs.  (6.15), (6.32) in \cite{BFHZ}, respectively,
up to an overall sign.  Corresponding to the energy result
(\ref{magnonenergy2}), we observe that
\be
p_{u}(u) = \chi_{2}(u) \,,
\label{magnonmomentum2}
\ee
where $\chi_{2}(u)$ is defined in (\ref{kernelKa}). The
$\vartheta$ particle also does not carry momentum.

\section{$S$ matrix}\label{sec:Smatrix}

We finally turn to the problem of computing the scattering amplitudes
for the various excitations.

\subsection{Spinon-spinon}

It is convenient to review the computation of the spinon-spinon $S$ matrix
\cite{ABL, Re} using the NLIE approach \cite{HRS}.
Let $\theta_{h_{1}}\,,
\theta_{h_{2}}$ denote the rapidities of the two spinons. Since
$\mathfrak{b}(\theta_{h_{1}} - i\epsilon)=-1$ (\ref{thetah}), the
$\ln  \mathfrak{b}$ equation  (\ref{bNLIE2}) implies
\be
i \pi =i D_{b}(\theta_{h_{1}}) +
\int_{-\infty}^{\infty}dx'\ K(\theta_{h_{1}}-x'-i/2) \ln Y(x') \,,
\label{lnbIR}
\ee
since the convolution terms involving $\mathfrak{B}$ and $\bar{\mathfrak{B}}$
become exponentially small in the IR limit.
Neglecting the convolution term in the $\ln y$ equation
(\ref{y1NLIE2}), one obtains
\be
y(x) = \tanh (\pi(x-\theta_{h_{1}})/2)
       \tanh (\pi(x-\theta_{h_{2}})/2) \,,
\label{ssy}
\ee
and therefore
\be
Y(x) = 1 + y(x) = {\cosh( \pi(x- (\theta_{h_{1}} +
\theta_{h_{2}})/2))\over \cosh (\pi(x-\theta_{h_{1}})/2)
\cosh (\pi(x-\theta_{h_{2}})/2)} \,.
\ee
We now exponentiate both sides of (\ref{lnbIR}), and note using
(\ref{Db1}), (\ref{Db2}) that
\be
D_{b}(\theta_{h_{1}}) = L \chi_{K}(\theta_{h_{1}}) +
\chi(\theta) -\pi\,, \qquad \theta = \theta_{h_{1}}
- \theta_{h_{2}} \,.
\ee
With the help of the momentum expression (\ref{spinonmomentum}),
we compare the result with the Yang equation
\be
e^{i L p_{h}(\theta_{h_{1}})}\ S_{h, h}(\theta) = 1 \,.
\ee
We conclude that the $S$ matrix is given (up to a constant) by
\be
S_{h, h}(\theta) = e^{i\chi(\theta)}
S_{RSOS}(\theta)  \,,
\label{ssSmatrix}
\ee
where
\be
S_{RSOS}(\theta)
&=& e^{\int_{-\infty}^{\infty}dx'\ K(\theta_{h_{1}}-x'-i/2) \ln
Y(x')} \non \\
&=& e^{-{i\over 2}\left[\psi_{0}(\theta) - \varphi_{2}(\theta)\right]}
= e^{-i\left[\psi_{0}(\theta) - \varphi_{4}(\theta)\right]}\,,
\label{RSOS}
\ee
and
\be
\psi_{0}(x) &=& \arctan \sinh (\pi x/2) =
i \ln{\sinh(\pi(i + x)/4)\over \sinh(\pi(i -x)/4)} \,, \label{psi0} \\
\varphi_{n}(x)  &=& \int_{0}^{\infty}dk\ {\sin(k x)
\sinh((n-1)k/2)\over k
\sinh(n k/2) \cosh(k/2)} \,, \label{varphin}
\ee
with $\varphi_{4}(x)=\left(\varphi_{2}(x) + \psi_{0}(x)\right)/2$.
The convolution integrals are performed using the results collected in
the appendix.  The result (\ref{RSOS}) is (up to a crossing factor,
and a rescaling of the rapidity by $\pi$)
one of the kink-kink scattering amplitudes of the tricritical Ising
model perturbed by the operator $\Phi_{(1,3)}$ \cite{RSOS}, which appears also in
the soliton-soliton $S$ matrix of the supersymmetric sine-Gordon
model \cite{ABL}. We note that
\be
\chi(\theta)= \frac{1}{i} \ln \frac{\Gamma(1 + i \theta/2)\, \Gamma(1/2- i \theta/2) }
{\Gamma(1 - i \theta/2)\, \Gamma(1/2 + i \theta/2)} \,,
\ee
which is (up to the same rescaling of the rapidity by $\pi$)
the soliton-soliton scattering phase of the sine-Gordon
model \cite{ZZ} in the isotropic limit $\beta^{2}\rightarrow 8\pi$.

If we also consider the $\ln y$ equation with an additional $i\pi$ term,
then the RHS of (\ref{ssy}) acquires a minus sign. The corresponding
amplitudes can be computed along similar lines \cite{HRS}. However,
for simplicity, we restrict our attention to the $\ln y$ equation
without this additional $i\pi$ term.

\subsection{Spinon-magnon}

Let $\theta_{h_{1}}\,, u_{1}$ denote the rapidities of the spinon and
magnon, respectively.
The spinon-magnon $S$ matrix can be computed in two different ways.
One way is to start from $\mathfrak{b}(\theta_{h_{1}} - i\epsilon)=-1$, which
again leads to (\ref{lnbIR}). The $\ln y$ equation implies
\be
y(x) = \tanh (\pi(x-\theta_{h_{1}})/2)
       \tanh (\pi(x-u_{1})/2) \,,
\ee
and therefore
\be
Y(x) = 1 + y(x) = {\cosh( \pi(x- (\theta_{h_{1}} + u_{1})/2))\over
\cosh (\pi(x-\theta_{h_{1}})/2) \cosh (\pi(x-u_{1})/2)} \,.
\label{smY}
\ee
Moreover, now $D_{b}(\theta_{h_{1}}) = L \chi_{K}(\theta_{h_{1}}) = L
p_{h}(\theta_{h_{1}})$, up to an additive constant.
Proceeding as before, we obtain the result
\be
S_{h, u}(\theta) = S_{RSOS}(\theta)  \,,
\label{smSmatrix}
\ee
where now $\theta= \theta_{h_{1}} - u_{1}$, and $S_{RSOS}(\theta)$ is given by
(\ref{RSOS}). That is, in contrast to the spinon-spinon $S$ matrix
(\ref{ssSmatrix}), the spinon-magnon $S$ matrix consists only of the RSOS factor.

A second way to compute the spinon-magnon $S$ matrix is to start from
$\mathfrak{c}(u_{1})=-1$ (\ref{uj}), which together with the $\ln
\mathfrak{c}$ equation (\ref{lnc}) imply
\be
i \pi = i D_{c}(u_{1}) +
\int_{-\infty}^{\infty}dx'\ K(u_{1}-x'-i/2) \ln Y(x') \,.
\label{lncIR}
\ee
We exponentiate both sides of this equation, and note that
\be
D_{c}(u_{1}) = L \chi_{2}(u_{1}) = L p_{u}(u_{1}) \,,
\label{Dcu1}
\ee
where we have made use of (\ref{Dc}) and the momentum result (\ref{magnonmomentum2}).
Comparing with the corresponding Yang equation, we recover the same
result, i.e.
\be
S_{u, h}(\theta) = S_{RSOS}(\theta)  \,,
\ee
where now $\theta= u_{1} - \theta_{h_{1}}$.

\subsection{Magnon-magnon}

Let $u_{1}$, $u_{2}$ be the rapidities of the two magnons.
The $\ln y$ equation (\ref{y1NLIE2}) implies
\be
y(x) = \tanh (\pi(x-u_{1})/2) \tanh (\pi(x-u_{2})/2)  \,,
\ee
and
\be
Y(x) = 1 + y(x) = {\cosh( \pi(x- (u_{1} + u_{2})/2))
\over \cosh (\pi(x-u_{1})/2) \cosh (\pi(x-u_{2})/2)} \,.
\ee
The condition $\mathfrak{c}(u_{1})=-1$ (\ref{uj}) and the $\ln
\mathfrak{c}$ equation (\ref{lnc}) again give (\ref{lncIR}),
where now (cf. (\ref{Dcu1}))
\be
D_{c}(u_{1}) =  L p_{u}(u_{1}) + \chi_{3/2}(\theta) \,,
\ee
with $\theta=u_{1}-u_{2}$. Proceeding as before, we conclude that the
magnon-magnon $S$ matrix is given by
\be
S_{u, u}(\theta) = e^{i \chi_{3/2}(\theta)}\,
S_{RSOS}(\theta) \,, \label{magnonmagnon1}
\ee
where $S_{RSOS}(\theta)$ is given by (\ref{RSOS}). We note that
\be
\chi_{3/2}(\theta)= \frac{1}{i}\ln
\frac{\Gamma(-1/2 + i \theta/2)\, \Gamma(-i \theta/2)}
{\Gamma(-1/2- i \theta/2)\, \Gamma(i \theta/2)} + \pi\,,
\label{magnonmagnon2}
\ee
and that $s(\theta) \equiv  e^{i \chi_{3/2}(\theta)}$ has the
crossing property
\be
s(i-\theta) = \left(\frac{1-i\theta}{2+i\theta}\right) s(\theta) \,.
\label{magnoncrossing}
\ee
Hence, $s(\theta)/(1+i\theta/2)$ is crossing invariant.

We have considered so far the composite operators containing only
$D_{\alpha{\dot 1}}$ covariant derivatives and computed the $S$ matrix
amplitude between them.  In principle, one would need to add $r$-roots
to compute amplitudes for the derivatives carrying the right-spin
state ${\dot 2}$.  But this can be done, without adding $r$-roots, by
using the $SU(2)_{R}$ symmetry.  The ``vertex'' part of the $S$
matrix is in fact a $4\times 4$ matrix which can be fixed completely
by the $SU(2)_{R}$ symmetry along with factorizability (i.e.,
Yang-Baxter equation), unitarity and crossing,
\be
\frac{s(\theta)}{1+i\theta/2}\left({\cal P} + i \theta/2 \right)\,,
\ee
where ${\cal P}$ is the permutation matrix.

\subsection{$\vartheta$-spinon and $\vartheta$-magnon}

The condition $\mathfrak{a}(\vartheta_{\alpha})=-1$ (\ref{varthetah})
together with the $\ln \mathfrak{a}$ equation (\ref{aNLIE1}) imply
that the $S$ matrices $S_{\vartheta, h}$ and $S_{\vartheta, u}$ are
identical, and are given by
\be
S(\theta) =
\frac{\sinh (\pi(\theta/2 - i/4))}
     {\sinh (\pi(\theta/2 + i/4))} \,.
\ee
The same result can also be obtained starting from
(\ref{thetah}), (\ref{bNLIE2}) (for $S_{h, \vartheta}$) and from
(\ref{uj}), (\ref{lnc}) (for $S_{u, \vartheta}$).  Since there is no
$\vartheta$-dependent contribution in the source term of the $\ln
\mathfrak{a}$ equation (\ref{aNLIE1}), there is no nontrivial
$\vartheta$-$\vartheta$ scattering.

\section{Discussion}\label{sec:conclude}

We have proposed a set of NLIEs (\ref{bNLIE2})-(\ref{Dy}),
(\ref{aNLIE1}), (\ref{Da}), (\ref{lnc})-(\ref{kernelKa}) to describe
the QCD spin chain of Beisert {\it et al.} \cite{BFHZ}.  We have used
these NLIEs to compute $S$ matrix elements for excitations of this
model, as shown in detail in Sec.  \ref{sec:Smatrix}.  The consistency
of our results ($S_{a, b}=S_{b, a}$ for particles $a$ and $b$ of
different types) provides further support for the validity of these
NLIEs.

Many questions remain to be addressed.  It should be possible to
generalize this work along the lines \cite{ANS} and compute the
boundary $S$ matrix for the open QCD spin chain corresponding to
operators with quarks at the ends.  The magnon-magnon $S$ matrix
(\ref{magnonmagnon1}), (\ref{magnonmagnon2}) has an infinite number of
singularities (starting at $\theta=\pm 2i$), which can presumably be
interpreted as magnon-magnon bound states (``breathers'').  The energy
and momentum of these breathers was computed using the string
hypothesis in \cite{BFHZ}.  It would be interesting to analyze these
excitations without invoking the string hypothesis, and to determine
their $S$ matrices.  It would also be interesting to consider the
effects of higher loops (\cite{FHZ} and \cite{BFHZ} worked only to
leading order in the `t Hoof coupling) and to better understand the
significance of these results for QCD, as well as for the full ${\cal
N}=4$ SYM theory and for the corresponding string theory.

\section*{Acknowledgments}
One of us (CA) thanks Shizuoka University and the University of Miami
for support.  This work was supported in part by
a Korea Research Foundation Grant funded by the Korean government
(MOEHRD) (KRF-2006-312-C00096) (CA),
by the National Science Foundation under Grants PHY-0244261 and
PHY-0554821 (RN),
and by the Ministry of Education of Japan, a Grant-in-Aid for
Scientific Research 17540354 (JS).

\begin{appendix}

\section{Convolutions}

The convolution integrals involving the kernel $K(x-{i\over
2})={i\over 2 \sinh \pi x}$ can be evaluated using the following
results
\be
{i\over 2}\int_{-\infty}^{\infty}dx'\ {\ln \cosh( \pi(x'-i\epsilon))\over
\sinh (\pi(x-x'+i\epsilon))} &=& -{i\over 2}\arctan \sinh (\pi x) +
{1\over 2}\ln \cosh(\pi x) \,, \\
{i\over 2}\int_{-\infty}^{\infty}dx'\ {\ln \cosh( \pi(x'-i\epsilon)/2)\over
\sinh (\pi(x-x'+i\epsilon))} &=& -{i\over 2}\varphi_{2}(x) +
{1\over 2}\ln \cosh(\pi x/2) \,, \\
{i\over 2}\int_{-\infty}^{\infty}dx'\ {\ln \sinh( \pi (x'-i\epsilon))\over
\sinh (\pi(x-x'+i\epsilon))} &=& {1\over 2}\ln \sinh (\pi x) -
{1\over 2}\ln \tanh(\pi x/2) -{i\pi\over 4}\,, \\
{i\over 2}\int_{-\infty}^{\infty}dx'\ {\ln \sinh( \pi(x'-i\epsilon)/2)\over
\sinh (\pi(x-x'+i\epsilon))} &=& {i\over 2}\varphi_{2}(x) +
{1\over 2}\ln \cosh(\pi x/2)  -{i\pi\over 4} \,,
\ee
where $\epsilon$ is a small positive number, and $\varphi_{2}(x)$ is
given by (\ref{varphin}).

\end{appendix}

\end{document}